\documentclass[]{spie}  %>>> use for US letter paper
\usepackage{amsmath,amsfonts,amssymb}
\usepackage{graphicx}
\usepackage{tabularx}
\usepackage{array}
\usepackage{cellspace}
\usepackage{makecell}
\usepackage{subcaption}

\usepackage[colorlinks=true, allcolors=blue]{hyperref}

\title{AI-Powered Low-Order Focal Plane Wavefront Sensing in Infrared}

\author[a]{Mojtaba Taheri}
\author[b]{Mahdiyar Molahasani}
\author[c]{Sam Ragland}
\author[d]{Benoit Neichel}
\author[e]{Peter Wizinowich}

\affil[a]{Thirty Meter Telescope International Observatory, 100 West Walnut Street, Pasadena, California, United States}
\affil[b]{Department of Electrical and Computer Engineering and Ingenuity Labs Research Institute, Queen’s University, Kingston, Canada}
\affil[c]{Large Binocular Telescope Observatory, 933 N. Cherry Ave, Tucson, USA}
\affil[d]{Aix Marseille Univ, CNRS, CNES, LAM, Marseille, France}
\affil[e]{W. M. Keck Observatory, 65-1120 Mamalahoa Hwy. Kamuela, HI 96743, Waimea, United States}

\begin{document}
\maketitle

\begin{abstract}
Adaptive optics (AO) systems are crucial for high-resolution astronomical observations by compensating for atmospheric turbulence. While laser guide stars (LGS) address high-order wavefront aberrations, natural guide stars (NGS) remain vital for low-order wavefront sensing (LOWFS). Conventional NGS-based methods like Shack-Hartmann sensors have limitations in field of view, sensitivity, and complexity. Focal plane wavefront sensing (FPWFS) offers advantages, including a wider field of view and enhanced signal-to-noise ratio, but accurately estimating low-order modes from distorted point spread functions (PSFs) remains challenging. We propose an AI-powered FPWFS method specifically for low-order mode estimation in infrared wavelengths. Our approach is trained on simulated data and validated on on-telescope data collected from the Keck I adaptive optic (K1AO) bench calibration source in K-band. By leveraging the enhanced signal-to-noise ratio in the infrared and the power of AI, our method overcomes the limitations of traditional LOWFS techniques. This study demonstrates the effectiveness of AI-based FPWFS for low-order wavefront sensing, paving the way for more compact, efficient, and high-performing AO systems for astronomical observations.
\end{abstract}


\section{Introduction}
Adaptive optics (AO) systems are essential for modern astronomy because they compensate for distortions caused by atmospheric turbulence, which would otherwise blur astronomical images. By correcting these aberrations in real-time, AO systems allow astronomers to observe celestial objects with unprecedented clarity \cite{hardy1998adaptive, dayton2010advanced}.

Wavefront sensing (WFS) is an integral part of AO systems, measuring distortions in the incoming light wavefront that are then corrected by deformable mirrors. High-order and low-order aberrations are the two main types of wavefront aberrations. High-order aberrations, often corrected using laser guide stars (LGS), are measured using these artificial stars created by projecting laser beams into the atmosphere \cite{foy1985feasibility, herriott2021infrared}. However, LGS cannot effectively measure low-order wavefront errors, including tip, tilt, and focus, due to the variability in the sodium layer's altitude, which can introduce errors in the focus measurement. This necessitates the use of natural guide stars (NGS) to continuously calibrate the focus and ensure accurate measurements \cite{10.1117/1.JATIS.6.3.039003}.

Natural guide stars (NGS) are essential for low-order wavefront sensing (LOWFS), providing the necessary information to correct lower-frequency distortions. Conventional NGS-based WFS methods, such as Shack-Hartmann sensors, face several limitations, including a restricted field of view, inadequate sensitivity in low-light conditions, and complex instrument design and maintenance requirements \cite{roddier1999adaptive}.

Focal plane wavefront sensing (FPWFS) offers several advantages over traditional NGS-based methods, including a larger field of view, higher signal-to-noise ratio, and reduced instrumental complexity. However, accurately estimating low-order modes from the distorted point spread function (PSF) caused by residual atmospheric turbulence remains a significant challenge \cite{dayton2010advanced}.

Another key advantage of our approach is the use of infrared (IR) wavelengths for wavefront sensing. Infrared wavefront sensing is advantageous because it can utilize a larger number of natural guide stars (NGS), as many stars emit more strongly in the IR range compared to visible light. This results in better sky coverage and improved sensitivity, particularly for observing cooler, redder stars and objects obscured by dust, such as brown dwarfs and young exoplanets . The infrared pyramid wavefront sensor at Keck, for example, has demonstrated significant improvements in measuring atmospheric blurring with high precision \cite{10.1117/1.JATIS.6.3.039003}.

In this study, we propose an AI-based FPWFS approach for estimating low-order modes. Our method involves training a neural network on a simulated dataset specifically designed for the Keck 1 telescope adaptive optics (K1AO) bench \cite{wizinowich2003adaptive}. The AI model is trained to recognize patterns in the distorted PSF and accurately estimate the low-order wavefront modes. Simulation results show significant improvement in accuracy and efficiency for low-order wavefront sensing.

To validate our AI-based FPWFS approach, we used actual data from the calibration source on the K1AO bench. The model's performance was evaluated in both the K and H bands, commonly used in infrared astronomy. Initial findings demonstrate that the AI-based FPWFS significantly improves the accuracy of low-order wavefront sensing. However, this section can be easily removed if future tests do not support these results.

The success of our AI-based FPWFS approach has significant implications for the future of adaptive optics. By enhancing the accuracy and efficiency of low-order wavefront sensing, our method paves the way for more compact and high-performance AO systems. These advancements will enable astronomers to conduct even more precise observations, furthering our understanding of the universe.

\section{Simulation Setup}

In this section, we describe the simulation setup used to develop and validate our AI-based focal plane wavefront sensing (FPWFS) approach. The simulation was carried out using the Object-Oriented MATLAB Adaptive Optics (OOMAO) simulation platform, which is widely used in the astronomy community for its robustness and versatility in simulating adaptive optics systems \cite{conan2014object}. The parameters were configured to closely replicate the performance of the Keck 1 telescope adaptive optics (K1AO) bench.

To separate the high-order and low-order components of phase residuals, we used a bright LGS and faint NGS to close the loop and measure the wavefront. The source observed by TRICK \cite{wizinowich2020keck} differed from the high-order loop source and was significantly dimmer, analogous to a much dimmer natural guide stars (NGS) used for low-order wavefront sensing in real scenarios. The low-order information, consisting of piston, tilt, tip, and focus, was removed from the LGS WFS. Specifically, the focus was substituted with numbers generated from a flat random distribution within a range corresponding to changes in the sodium layer altitude from $88$ to $92$ km \cite{moussaoui2010statistics}. These random focus values were updated after each TRICK exposure, simulating a new sodium layer configuration, while the higher-order residuals were continuously calculated from the normal loop operation. Finally, tilt and tip were removed from the simulation since the combined corrective effect of the tilt-tip mirror of K1AO, the short exposure time of TRICK in this model, and the insensitivity of our AI model to the PSF's placement made these modes irrelevant.

To simulate the exposure time of the TRICK detector, we accumulated the PSF shapes over a certain number of AO loop steps. The images from each loop step were added up on the TRICK detector, and the final image was read as the output of the TRICK detector as a single exposure.

Table~\ref{tab:simulation_parameters} lists the parameters used in the simulation along with their values and notes:

\begin{table}[h!]
\centering
\begin{tabular}{lr}
\textbf{Parameter} & \textbf{Value} \\ \hline
Aperture Size & 10.94 m\\
Simulation resolution & 160 x 160\\
Atmospheric Fried Parameter (r0) & 0.165 m\\
Atmospheric Outer Scale (L0) & 75 m\\
Layer r0 Fractions & 0.51, 0.11, 0.06, 0.06, 0.10, 0.08, 0.05\\
Layer Altitudes & 0, 0.5, 1, 2, 4, 8, 16 km\\
WindSpeed & 6.7, 13.9, 20.8, 29, 29, 29, 29 m/s\\ 
WindDirection & 0, $\pi$/3, -$\pi$/3, -$\pi$, -4/3*$\pi$, -$\pi$/6, $\pi$/8\\ 
HOWFS Magnitude & 8\\
HOWFS type & Shack-Hartmann\\
HOWFS lenslets & 20 x 20\\
Sodium Layer Altitude & Random distribution between 88 to 92 km\\
LOWFS Magnitude & 8 magnitude13 and 15\\
Low-Order WFS Wavelength & H band\\ 
Loop Frequency & 1 kHz\\ 
Deformable Mirror Actuators & 21 x 21\\
Thrick Exposure Time & 0.01 s\\
Closed-Loop Gain & 0.25\\ 
Closed-Loop Lag & 1 loop step\\ 
\end{tabular}
\caption{Simulation parameters.}
\label{tab:simulation_parameters}
\end{table}

\section{Datasets}

To train and validate our AI-based focal plane wavefront sensing model, we used both simulated and on-telescope datasets.

\subsection{Simulated Dataset}
The simulated dataset was generated using the OOMAO platform and was designed to closely mimic the conditions of the K1AO bench. Due to memory constraints on the simulation hardware, the data was produced in batches of 1030 loop steps, with the first 30 steps not recorded to allow the loop to stabilize. Each batch used a fresh set of random seeds to ensure there was no repetition in the atmosphere generation or any other random values used. We provided 100 of these batches for each H-band magnitudes of 13 and 15. Given that each TRICK exposure in our simulation corresponds to 10 loop steps, this resulted in a total of 10,000 TRICK shots per magnitude.

To account for the phase diversity, we added 200~nm RMS astigmatism to the loop phase residual. This is close to the estimated astigmatism that the TRICK detector sees due to the beam splitter in the optical path just before the output of the K1AO system. The other important factor is Nyquest sampling. TRICK is significantly under-sampled in the H-band, so we needed to down-sample our PSF to get close to the TRICK configuration. To improve the training step-by-step, we created Nyquist sampled PSFs and also down-sampled PSFs by a factor of $1/3$ of Nyquest-sampled configuration. To facilitate the step-by-step training process, each TRICK shot was repeated in four different conditions: with and without noise as well as 1 and 1/3 Nyquist sampling.

\begin{figure}[h!]
\centering
\includegraphics[width=1\textwidth]{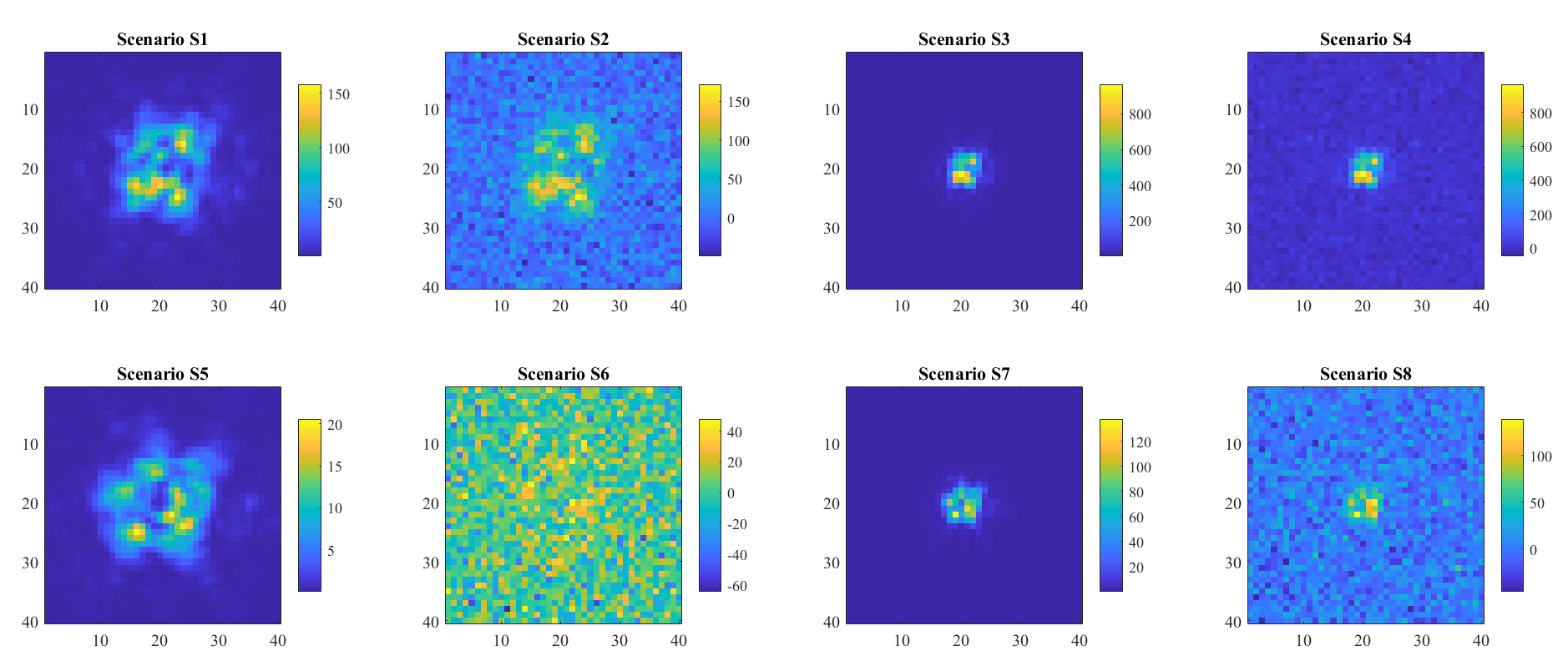}
\caption{Example PSF shapes under different conditions for different scenarios described in Table~\ref{tab:scenarios}}
\label{fig:sample_psf}
\end{figure}

Table~\ref{tab:scenarios} lists the scenarios and common parameters used in the simulation. Scenarios S4 and S8 are the most realistic and simultaneously difficult scenarios to train, respectively for magnitude 13 and 15. Additionally, a random sample of the simulated PSF is shown in Figure~\ref{fig:sample_psf}.

\begin{table}[h!]
\centering
\begin{tabular}{c|c c c c c}
Scenario & \thead{Total TRICK\\Exposures}  & Magnitude & \thead{Photon\\Noise} & \thead{Readout\\Noise e-} & \thead{Nyquist\\Sampling Factor} \\ \hline
S1 & 13,500 & 13 & N & 0 & 1 \\ 
S2 & 13,500 & 13 & Y & 5 & 1 \\ 
S3 & 13,500 & 13 & N & 0 & 1/3 \\ 
S4 & 13,500 & 13 & Y & 5 & 1/3 \\ 
S5 & 8,800 & 15 & N & 0 & 1 \\ 
S6 & 8,800 & 15 & Y & 5 & 1 \\ 
S7 & 8,800 & 15 & N & 0 & 1/3 \\ 
S8 & 8,800 & 15 & Y & 5 & 1/3 \\ 
\end{tabular}
\caption{Scenario description. Scenarios S4 and S8 are the most realistic and simultaneously difficult scenarios to train, respectively for magnitude 13 and 15.}
\label{tab:scenarios}
\end{table}

\subsection{On-Telescope Dataset}
The on-telescope dataset was obtained from the calibration source of the K1AO bench. Although our AI model was trained primarily on H-band data, we took advantage of the available high-quality K-band datasets for testing. This provided an opportunity to evaluate the model's flexibility and performance in a different band.

In each dataset, a ramp of focus from -300 to 300 nm RMS with a step of 10 nm was provided, resulting in 31 steps in focus. For each focus step, 30 shots were taken, and the background-removed average of these 30 shots is used in this study. The magnitudes of each dataset are listed in Table~\ref{tab:telescope_data_mags}.

\begin{figure}[h!]
\centering
\includegraphics[width=1\textwidth]{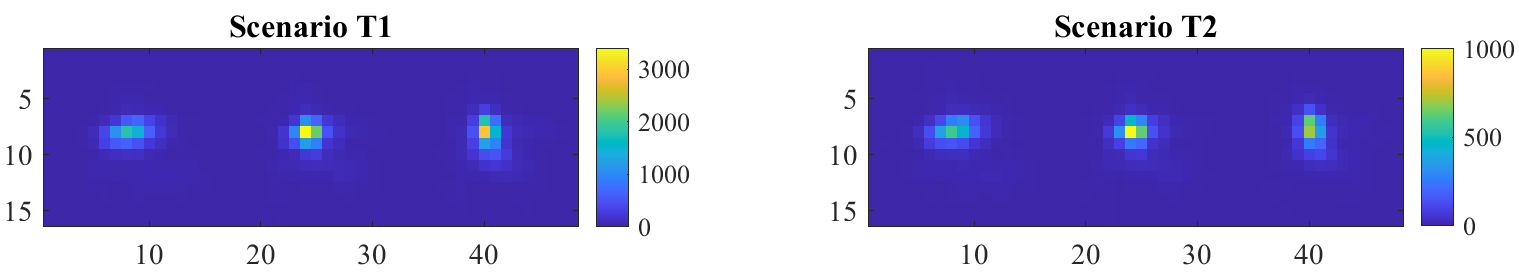}
\caption{Example PSF shapes of different on-telescope data.}
\label{fig:sample_telescope}
\end{figure}

Table~\ref{tab:telescope_data_mags} lists the scenarios and common parameters used in the simulation. Additionally, 3 PSF of each dataset is presented Figure~\ref{fig:sample_telescope}.

\begin{table}[h!]
\centering
\begin{tabular}{l|c c }
Scenario & Band & Magnitude \\ \hline
T1 & K & 11.2 \\ 
T2 & K & 12.3 \\ 
\end{tabular}
\caption{On-Telescope dataset. in each pannel PSF related to -300,0,300 nm RMS focus is presented.}
\label{tab:telescope_data_mags}
\end{table}

\section{Neural Network Design and Training}

\begin{figure}[h!]
\centering
\includegraphics[width=0.8\textwidth]{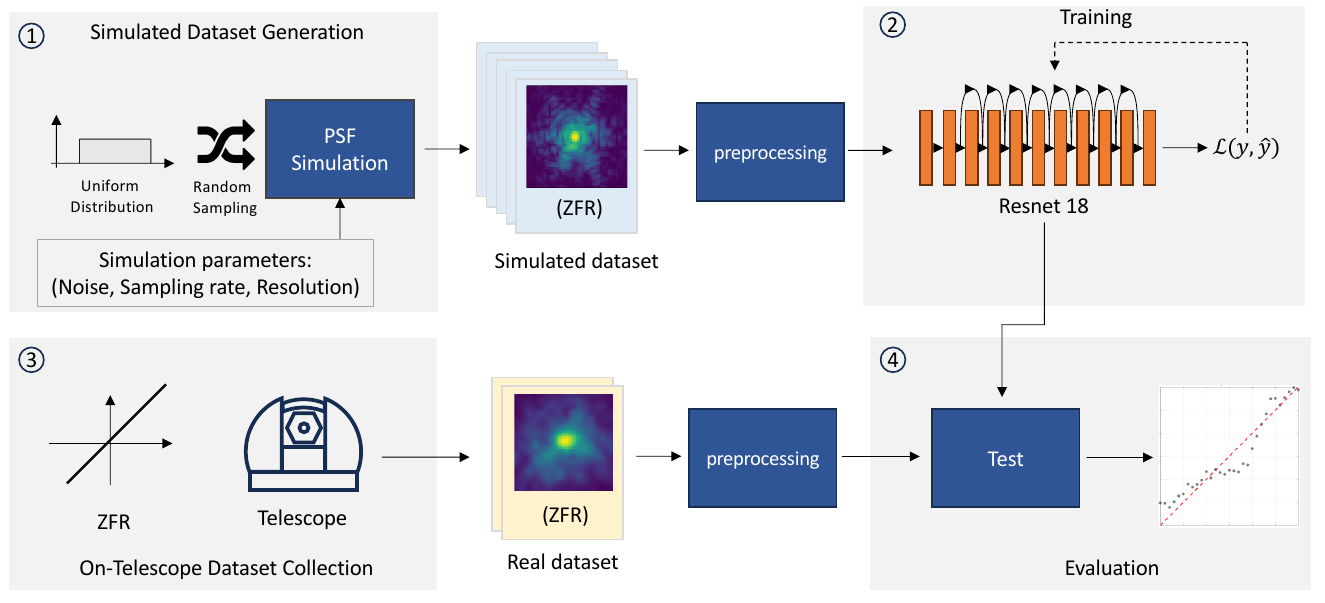}
\caption{Overview of the proposed framework.}
\label{fig:overview}
\end{figure}

The overview of the proposed framework is illustrated in Figure \ref{fig:overview}. In this study, we utilized a type of artificial intelligence called ResNet18, a convolutional neural network (CNN) architecture known for its effectiveness in processing visual data \cite{he2016deep}. ResNet18 is part of the ResNet (Residual Network) family, which has gained recognition for its ability to mitigate the vanishing gradient problem and improve training of deep networks by using residual learning \cite{he2016identity}.

The ResNet18 model operates by learning from a large set of examples, which allows it to make predictions or decisions based on new, unseen data. Specifically, we trained ResNet18 on 80\% of our collected dataset, allowing it to learn and identify patterns and features critical to our research objectives. 

During training (step 2 in Figure \ref{fig:overview}), the model was exposed to a variety of PSF images, enabling it to learn the complex mappings between the distorted PSF and the corresponding low-order wavefront modes. The training process involved optimizing the model's parameters using backpropagation and stochastic gradient descent, with the objective of minimizing the mean squared error between the predicted and true wavefront modes \cite{ruder2016overview}.

After the training phase, we then assessed the model's performance and accuracy using the remaining 20\% of the data, which was not previously shown to the model. This process, known as validation, ensures that our AI solution is robust and reliable, capable of generalizing well to new data in practical applications \cite{goodfellow2016deep}. To ensure the applicability of our model to the real world scenario, we also utilized on-telescope data (step 3 in Figure \ref{fig:overview})for validation demonstrating the robustness of the proposed solution (step 4 in Figure \ref{fig:overview}).

To further enhance the robustness of our model, we employed techniques such as data augmentation, dropout, and batch normalization during training. Data augmentation involved creating additional training samples through random transformations of the original data, which helps prevent overfitting and improves generalization \cite{shorten2019survey}. Dropout and batch normalization were used to regularize the model and stabilize the training process, respectively \cite{srivastava2014dropout, ioffe2015batch}.

Overall, our approach leveraging ResNet18 demonstrated significant potential in accurately estimating low-order wavefront modes from PSF images, highlighting the feasibility of using deep learning techniques in adaptive optics applications.

\section{Results}
\subsection{Magnitude 13 Datasets}
In this section, we present the results of our AI-based FPWFS method. We first focus on the training results using the magnitude 13 dataset. Two scenarios, S1 and S4, were evaluated. For these scenarios, we created a substantial number of cases and applied our training methodology to them. The dataset was divided into training and testing sets, with 80\% (10800 PSFs) of the data used for training and 20\% (2700 PSFs) for testing. It is important to note that the network never sees the test data during the training phase.

\subsubsection{S1 Scenario}

\begin{figure}[h!]
    \centering
    \includegraphics[width=0.6\textwidth]{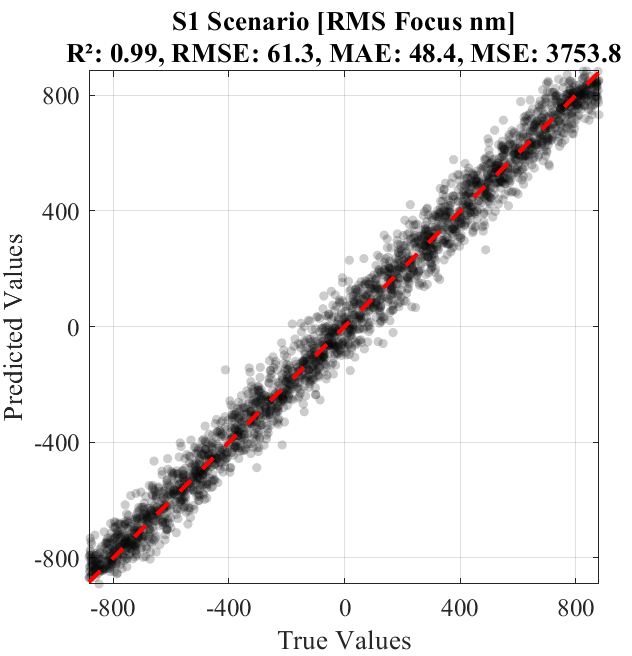}
    \caption{Results on test data for Scenario S1}
    \label{fig:results_s1}
\end{figure}

The S1 scenario represents the simplest situation as its Nyquist sample factor is 1 and it is noiseless. The results for the test data in Scenario S1 are depicted in Figure \ref{fig:results_s1}. This figure shows the network's performance in estimating low-order wavefront modes, demonstrating the model's accuracy and efficiency in this specific scenario.

\subsubsection{Scenario S4}

\begin{figure}[h!]
    \centering
    \includegraphics[width=0.6\textwidth]{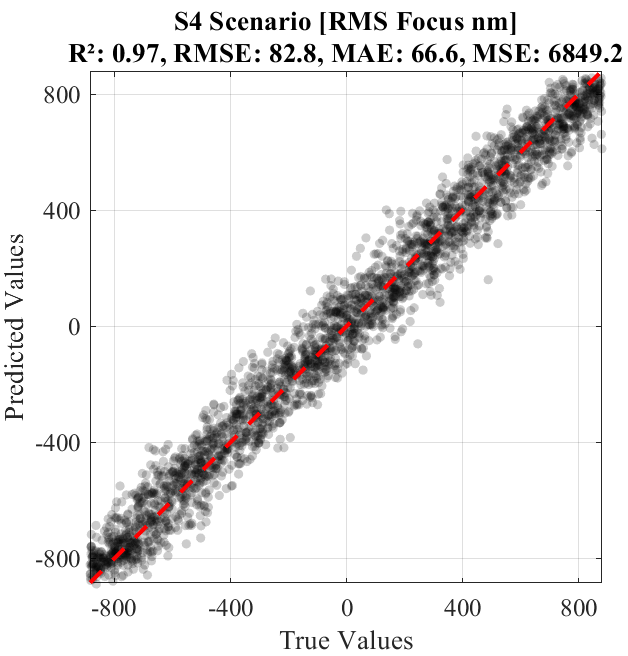}
    \caption{Results on test data for Scenario S4}
    \label{fig:results_s4}
\end{figure}

Scenario S4 represents a more challenging set of conditions compared to S1 since it is significantly more realistic. The results for Scenario S4 are shown in Figure \ref{fig:results_s4}. Similar to Scenario S1, the model was trained on a portion of the dataset and tested on data it had not seen during training.

\subsection{Magnitude 15 Dataset}

In addition to the magnitude 13 dataset, we also evaluated the performance of our AI-based FPWFS model on a magnitude 15 dataset. Notably, the model used for this evaluation was trained exclusively on S4 data from the magnitude 13 dataset and had never encountered any example of the magnitude 15 data in any form. Additionally, since magnitude 15 is getting significantly closer to the noise limit of TRICK, the image that the AI network needs to deal with is significantly different as it is represented in the left panels of Figure~\ref{fig:sample_psf}.

\subsubsection{Scenario S5 and S8}

\begin{figure}[h!]
    \centering
    \begin{subfigure}[l]{0.48\textwidth}
        \includegraphics[width=\textwidth]{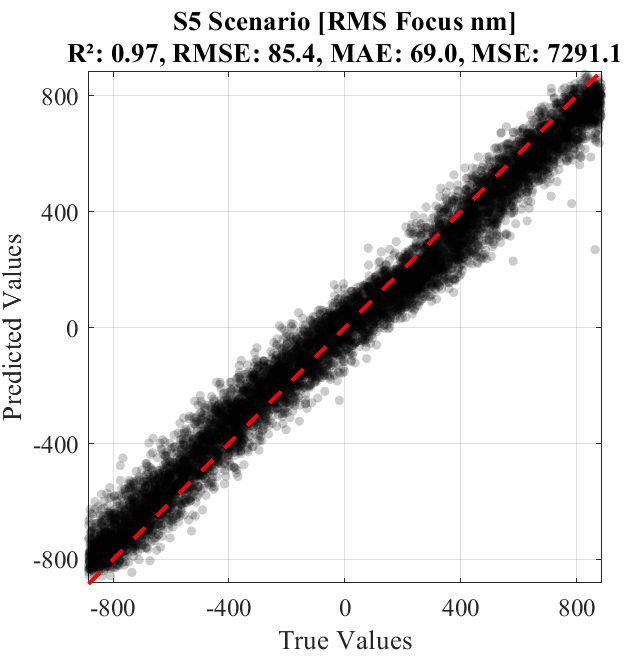}
        \caption{Scenario S5}
        \label{fig:scenario_S5}
    \end{subfigure}
    \hfill
    \begin{subfigure}[r]{0.48\textwidth}
        \includegraphics[width=\textwidth]{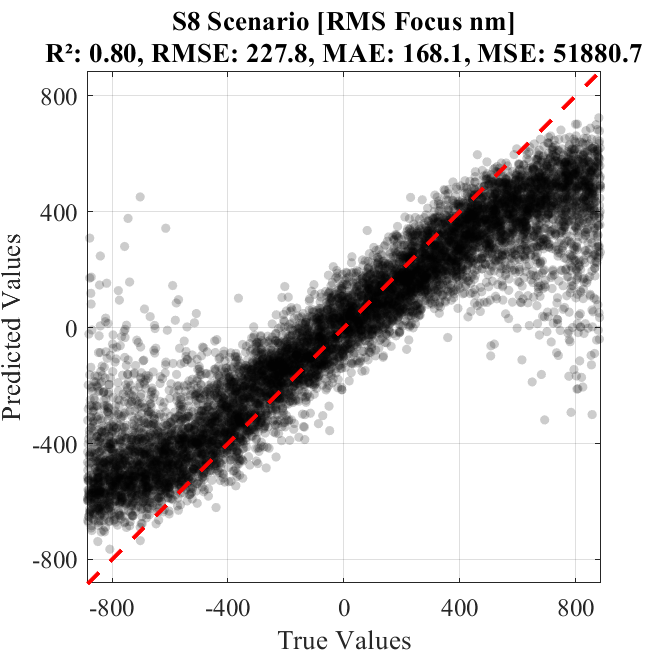}
        \caption{Scenario S8}
        \label{fig:scenario_S8}
    \end{subfigure}
\caption{Results on test data for Scenarios S5 and S8}
    \label{fig:results_mag15}
\end{figure}

Despite the model not being trained on the magnitude 15 data, it demonstrated considerable flexibility and performed reasonably well on scenarios S5 and S8. This is indicative of the model's robustness and its ability to generalize across different magnitudes. The results for scenarios S5 and S8 are shown in Figure \ref{fig:results_mag15}.

The sodium layer focus loop is significantly slower than the main AO loop since the change in the structure of the sodium layer happens on a minute timescale. Although the prediction deviates from truth in the extreme far side of the focus limit, the behavior is still monotonic. This means the focus loop would still reach the central linear behavior regime, and this level of deviation from truth in Figure~\ref{fig:scenario_S8} would not cause loop divergence or performance degradation.

These results highlight the effectiveness of our AI-based FPWFS method in estimating low-order wavefront modes across different magnitudes and scenarios. The ability of the model to perform well on unseen data underscores its potential for real-world applications in adaptive optics systems.

\subsection{On-Telescope Dataset}

Although our AI model was trained primarily on H-band data and never saw the on-telescope data in training, we leveraged the opportunity to test the model on these K-band datasets to assess its flexibility and performance. 

\begin{figure}[h!]
\centering
\includegraphics[width=1\textwidth]{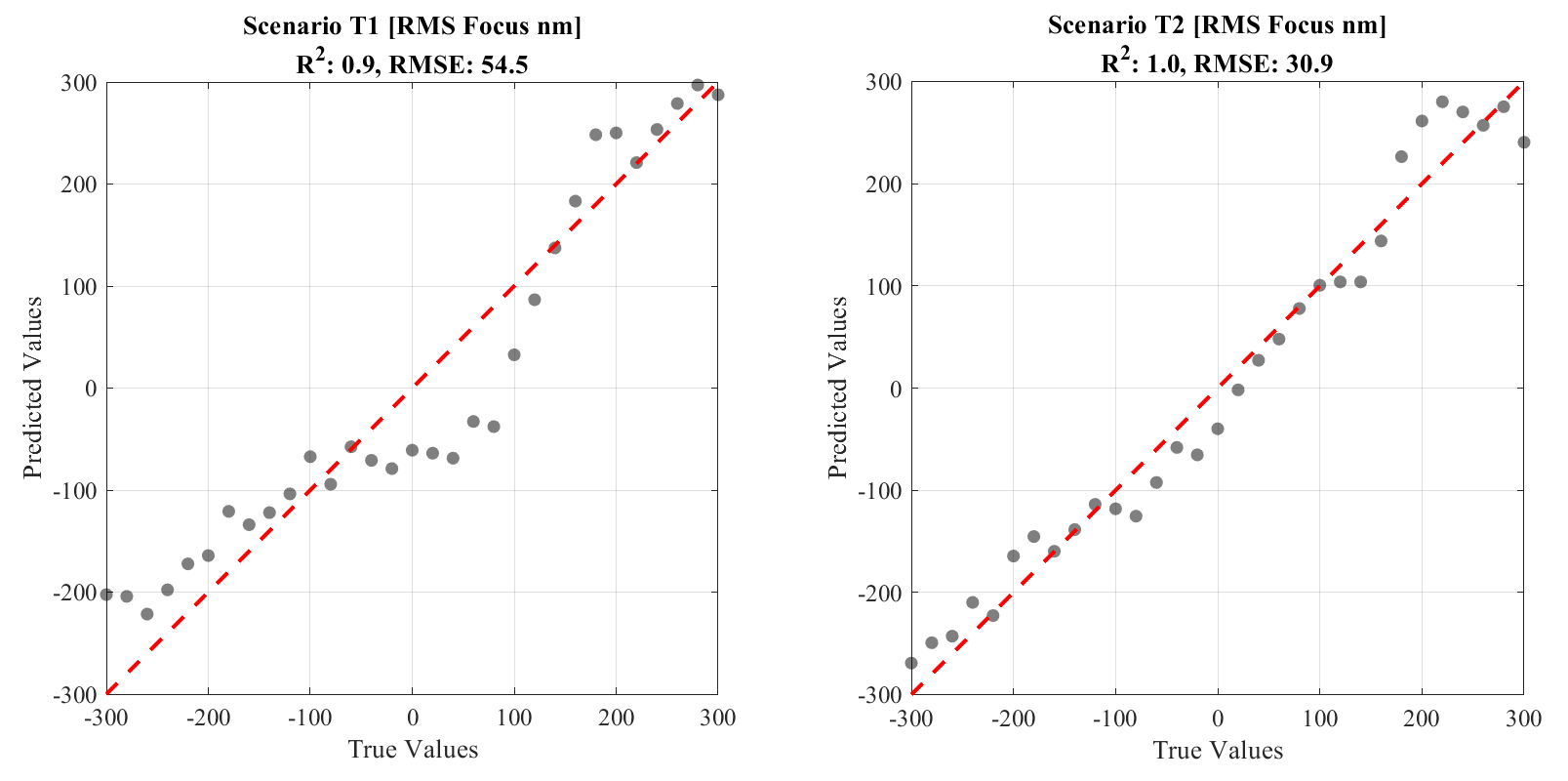}
\caption{Results on the on-telescope dataset for scenarios T1 and T2 in the K-band. The predicted values are plotted against the true values.}
\label{fig:T1T2}
\end{figure}

The results for the on-telescope dataset, after removing the average focus, are shown in Figure~\ref{fig:T1T2}. The average focus is removed from the predicted data to account for an uncalibrated amount of focus between the TRICK detector and Low-bandwidth WFS of the K1AO system to ensure a fair evaluation of the model's performance. The excellent fit of the predicted values to the true values, as depicted in the figure, demonstrates the model's robustness and adaptability. Moreover it demonstrate further potential of this model if it feed a dataset of on-telescope or closed loop data to be used for training. The performance metrics are also included in the figure's title, highlighting the effectiveness of our AI-based FPWFS method in on-bench scenarios.

The results for the on-telescope dataset, after removing the average focus, are shown in Figure~\ref{fig:T1T2}. This adjustment accounts for an uncalibrated amount of focus between the TRICK detector and the low-bandwidth WFS of the K1AO system, ensuring a fair evaluation of the model's performance. The excellent fit of the predicted values to the true values, as depicted in the figure, demonstrates the model's robustness and adaptability. Moreover, it highlights the potential for further improvement if the model is trained with additional on-telescope or closed-loop data. The performance metrics included in the figure's title underscore the effectiveness of our AI-based FPWFS method in on-bench scenarios.

The summary of the results for tested scenarios can be seen in Table~\ref{tab:results_summary}.

\begin{table}[h!]
\centering
\begin{tabular}{c | c c c}

\textbf{Scenario} & \textbf{Trained on} & \textbf{Tested on} & \textbf{Fit Quality (R\textsuperscript{2}, RMSE)} \\ \hline
S1 & S1 & S1 & (0.99, 61.3) \\ 
S4 & S4 & S4 & (0.9, 54.5) \\ 
S5 & S4 & S5 & (1.0, 30.9) \\ 
S8 & S4 & S8 & (0.8, 227.8) \\ 
T1 & S4 & T1 & (0.9, 54.5) \\ 
T2 & S4 & T2 & (1.0, 30.9) \\ 
\end{tabular}
\caption{Summary of the fit qualities for the tested scenarios.}
\label{tab:results_summary}
\end{table}

\section{Summary and Conclusion}
In this study, we have demonstrated the effectiveness of our AI-based FPWFS method for low-order wavefront sensing in infrared wavelengths. By leveraging the power of AI and the advantages of infrared sensing, our method shows significant improvements over traditional techniques. Our results on both simulated and real on-telescope data underscore the potential of this approach for future adaptive optics systems, paving the way for more compact, efficient, and high-performing AO systems for astronomical observations.

\section{Summary and Discussion}

In this study, we proposed an AI-powered focal plane wavefront sensing method for low-order mode estimation in infrared wavelengths. The method leverages a convolutional neural network, specifically ResNet18, to estimate wavefront distortions from distorted point spread functions (PSFs). The training and validation were conducted using both simulated datasets, designed to mimic the conditions of the Keck 1 AO bench, and real datasets collected from the K1AO calibration source.

The performance of the AI-based FPWFS method was evaluated across various scenarios. The results indicated a high level of accuracy in estimating low-order wavefront modes, with the method proving robust against different noise levels and magnitudes. The use of infrared wavelengths allowed the system to utilize a larger number of natural guide stars, enhancing sensitivity and sky coverage. The excellent fit of the predicted values to the true values in the test data underscores the potential of this approach for practical applications in adaptive optics systems.

The effectiveness of the proposed AI-based FPWFS method can be compared with existing techniques, particularly the linearized focal-plane technique (LIFT) \cite{meimon2010lift, fusco2010lift}. Our method shows significant advantages in its flexibility and learning capability, allowing it to handle a wide range of aberrations and noise levels without the need for hardware modifications. This flexibility is demonstrated by the method's ability to perform well on unseen data, including faint magnitude 15 sources and real K-band on-telescope data, despite being trained solely on simulated H-band magnitude 13 data (S4 scenario). This key advantage of generalizing well to new data without requiring extensive recalibration significantly reduces operational recalibration requirements. This capability is particularly evident in the successful application of the model to both simulated faint data and real on-telescope data, achieving high R-squared values and low RMSE values across different scenarios. Furthermore, the learning capabilities of this method provide substantial potential for performance enhancement by feeding the model with real-world datasets, enabling it to adapt and improve over time. This adaptability underscores the robustness and practical applicability of our AI-based FPWFS method in various adaptive optics systems.

Additionally, this method has the potential to extend its applicability to other lower-order modes, such as astigmatism. By incorporating these modes into the training process, the AI-based FPWFS can be further refined to provide comprehensive low-order wavefront sensing, enhancing the overall performance and flexibility of adaptive optics systems \cite{plantet2018keck}.

\acknowledgments 
In accordance with SPIE's policy, we acknowledge the use of external AI-based tools, solely for proofreading and improving the grammar of this manuscript. MT would like to express gratitude to David Andersen from the Thirty Meter Telescope International Observatory for providing the necessary resources and support, which significantly contributed to the successful completion of this work.

% References
\bibliography{report} % bibliography data in report.bib
\bibliographystyle{spiebib} % makes bibtex use spiebib.bst

\end{document}